\documentclass[preprint,showpacs,preprintnumbers,amsmath,amssymb]{revtex4}

\usepackage{graphicx}% Include figure files
\input{axodraw.sty}
\input{epsf}
\usepackage{rotate}
\def \beq{\begin{equation}}

\def \eeq{\end{equation}}

\def \beqa{\begin{eqnarray}}
\def \eeqa{\end{eqnarray}}
\begin{document}
\mark{{}{Y.-Y. Keum}}
\title{Phenomenological Applications of $k_T$ factorization}

\author{Yong-Yeon Keum\footnote{Email:yykeum@phys.sinica.edu.tw;yykeum@mail.desy.de}}
\affiliation{Institute of Physics, Academia Sinica, Nankang 128, Taipei 11529, Taiwan, \\
Deutsches Elektronen-Synchrotron DESY, 22607 Hamburg, Germany}

\begin{abstract}
We discuss applications of the perturbative QCD approach in the exclusive
non-leptonic two body B-meson decays.
We briefly review its ingredients and some important theoretical
issues on the factorization approach. 
PQCD results are compatible with present experimantal data for charmless
B-meson decays. We predict the possibility of large direct CP
asymmetry in $B^0 \to \pi^{+}\pi^{-}$ $(23\pm7 \%)$ and
$B^0\to K^{+}\pi^{-}$ $(-17\pm5\%)$. 
We also investigate the Branching ratios, CP asymmetry and isopsin symmetry
breaking in radiative $B \to (K^*/\rho) \gamma$ decays.
\end{abstract}

\keywords{Nonleptonic B-decays, $k_T$ factorization, Large Direct CP-violation}

%\pacs{2.0}
\maketitle

\section{Introduction}
The aim of the study on weak decay in B-meson is two folds:
(1) To determine precisely the elements of 
Cabibbo-Kobayashi-Maskawa (CKM) matrix\cite{Cabibbo,KM} and 
to explore the origin of CP-violation at a low energy scale,
(2) To understand strong interaction physics related to the confinement
of quarks and gluons within hadrons.

The two tasks complement each other. An understanding
of the connection between quarks and hadron properties is 
a necessary prerequeste for a precise determination of CKM matrix
elements and CP-violating phases, 
so called Kobayashi-Maskawa(KM) phase\cite{KM}. 

The theoretical description of hadronic weak decays is difficult
since nonperturbative QCD interactions is involved. This makes  
a difficult to interpret correctly data from asymmetric B-factories
and to seek the origin of CP violation. In the case of B-meson decays
into two light mesons, we can explain roughly branching ratios by 
using the factorization approximation \cite{BSW:85,BSW:87}.
Since B-meson is quite heavy, when it decays into two light mesons,
the final-state mesons are moving so fast that it is difficult to
exchange gluons between final-state mesons. So we can express the
amplitude in terms of the product of weak decay constant 
and transition form factors 
by the factorization (color-transparancy) argument\cite{Bro,Bej}. 
In this approach we neglect non-factorizable
contributions and a power suppressed annihilation contributions. 
Because of this weakness, asymmetry
of CP violation can not be predicted correctly.

Recently two different QCD
approaches beyond naive and general factorization assumption
\cite{BSW:85,BSW:87,Ali:98,Cheng:99} 
was proposed:
(1) QCD-factorization in the heavy quark limit \cite{BBNS:99,BBNS:00}
 in which non-factorizable terms and $a_{i}$ are calculable in some cases. 
(2) A Novel PQCD approach \cite{KLS:01,KLS:02,KLS:03} including 
the resummation effects of the transverse momentum carried by partons
inside meson.
In this review paper, we discuss some important theoretical issues in the PQCD
factorization and numerical results for charmless B-decays at the section 
3-7. In section 8 we present the PQCD results 
of the radiative B-decays $B \to K^{*}\gamma, 
\rho\gamma$.

\section{$k_T$ Factorization vs Collinear Factorization}
Let`s start to review shortly on the developement of theoretical
methods of exclusive hadronic B-meson two-body decays,
and make a comparision between different frameworks of factorization.

The theoretical basis how we can explain nonleptonic B-meson decays is
origined from the color-transparancy argument\cite{Bro,Bej}:
{\it Since b-quark decays into light quarks energetically ($ > 1$ GeV), 
the produced quark-antiquark pair doesn`t have enough time to evolve
to the real size hadronic entity, but remains a small size bound state
with a correspondingly small chromomagnetic moment which suppress in QCD
interaction between final state mesons.} 
%%%%%%%%%%%%%%%%%%%%%%%%%%
\vskip0.5cm
\begin{figure}[ht]
\epsfxsize=10cm
\centerline{\epsfbox{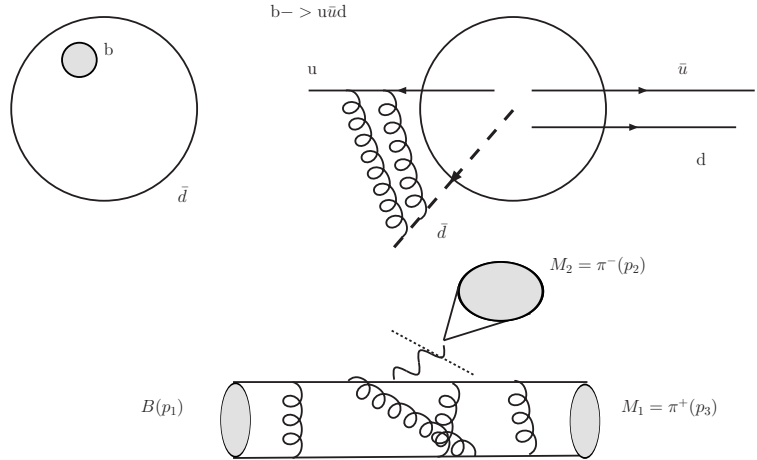}}
\begin{center}
\caption{Color transparancy arguement of the nonleptonic B-decays.}
\end{center}
\label{fig:color1}
\end{figure}
%%%%%%%%%%%%%%%%%%%%%%%%%%%  
As shown in Fig.~1, when we consider B-meson decays into
two final state pions,
the matrix elements of $0_1$ can be expressed by a simple way
as so called the naive factorization method\cite{BSW:85,BSW:87}:
\begin{eqnarray}
<\pi(p_2)\pi(p_3)|0_1(\mu)|B(p_1)> &\sim&
<\pi(p_2)|(d_i q_i)_{V-A}|0>
<\pi(p_3)|(q_j b_j)_{V-A}|B(p_1)> \nonumber \\
&=& \hspace{20mm} f_{\pi}
\hspace{10mm} \otimes \hspace{10mm} F^{B\pi}(q^2=M_{\pi}^2)
\end{eqnarray} 
In this way, only factorizable part was considered, but not nonfactorizable part.
In 1996, the generalied factorization approach was developped by A. Kamal\cite{Kamal}
and H.-Y. Cheng\cite{HYCheng}, which included non-factorizable contributions into
the effective wilson coefficients by assuming $NF = \chi \otimes F$.
The generalized factorization has a weak point to predict the CP asymmetry, 
since they considered non-factorizable part to be real, 
however it is complex in general.  
After then, QCD-factorization was proposed by Beneke et al.\cite{BBNS:99},
which is an improved form of naive factorization approach.
When we consider $B \to M_1 M_2$ with recoiled $M_1$ and emitted 
$M_2$(light or quarkonium), soft gluon exchanged effects are confined to
($BM_1$) system and only hard interactions between ($BM_1$) and $M_2$ survive 
in $m_b \to \infty$ limit which is calculable perturbatively.
The decay amplitude can be written as:
\begin{equation}
{\rm Decay\,\,Amp} = {\rm Amp}_{(naive \,\, fact)} \otimes [ 1 + O(\alpha_s) 
+O({\Lambda_{QCD} \over m_b})].
\end{equation}
In principle, nonfactorizable part and $a_i$ are calculable in the heavy quark limit
within the leading twist, 
however it is diverged at the end point with the twist-3 contributions
and even the leading twist contribution in the annihilation diagram.
To solve this end-point sigularity problem, PQCD approach was proposed by
Keum et al.\cite{KLS:01}, in which hard gluon exchanged contributions is dominant
even in the ($BM_1$) transition form factor. 
We will discuss the detail of PQCD approach in the next section. \\

Now we explain how to derive collinear and $k_T$ factorization
theorems for the pion form factor involved in the scattering process
$\pi(P_1)\gamma^*(q)\to\pi(P_2)$. The momenta are chosen in the 
light-cone coordinates as $P_1=(P_1^+,0,{\bf 0}_T)$,
$P_2=(0,P_2^-,{\bf 0}_T)$, and $Q^2=-q^2$.
At leading order, $O(\alpha_s)$, shown
in Fig.~2(a), the hard kernel is proportional to $H^{(0)}(x_1,x_2)\propto 
-1/(x_1P_1-x_2P_2)^2=1/(x_1x_2Q^2)$. Here $x_1$ and $x_2$ are the parton
momentum fractions carried by the lower quarks in the incoming and
outgoing pions, respectively.
At next-to-leading order, $O(\alpha_s^2)$, collinear divergences are 
generated in loop integrals, and need to be factorized into the 
pion wave function. In the collinear region
with the loop momentum $l$ parallel to 
$P_1$, we have an on-shell gluon 
$l^2\sim P_1^2\sim O(\Lambda^2)$ with the hierachy of the components,
$l^+\sim P_1^+ \gg l_T\sim \Lambda 
\gg l^-\sim \Lambda^2/P_1^+$.
%%%%%%%%%%%%%%%%%%%%%%%%%%
\vskip0.3cm
\begin{figure}[ht]
\epsfxsize=8cm
\centerline{\epsfbox{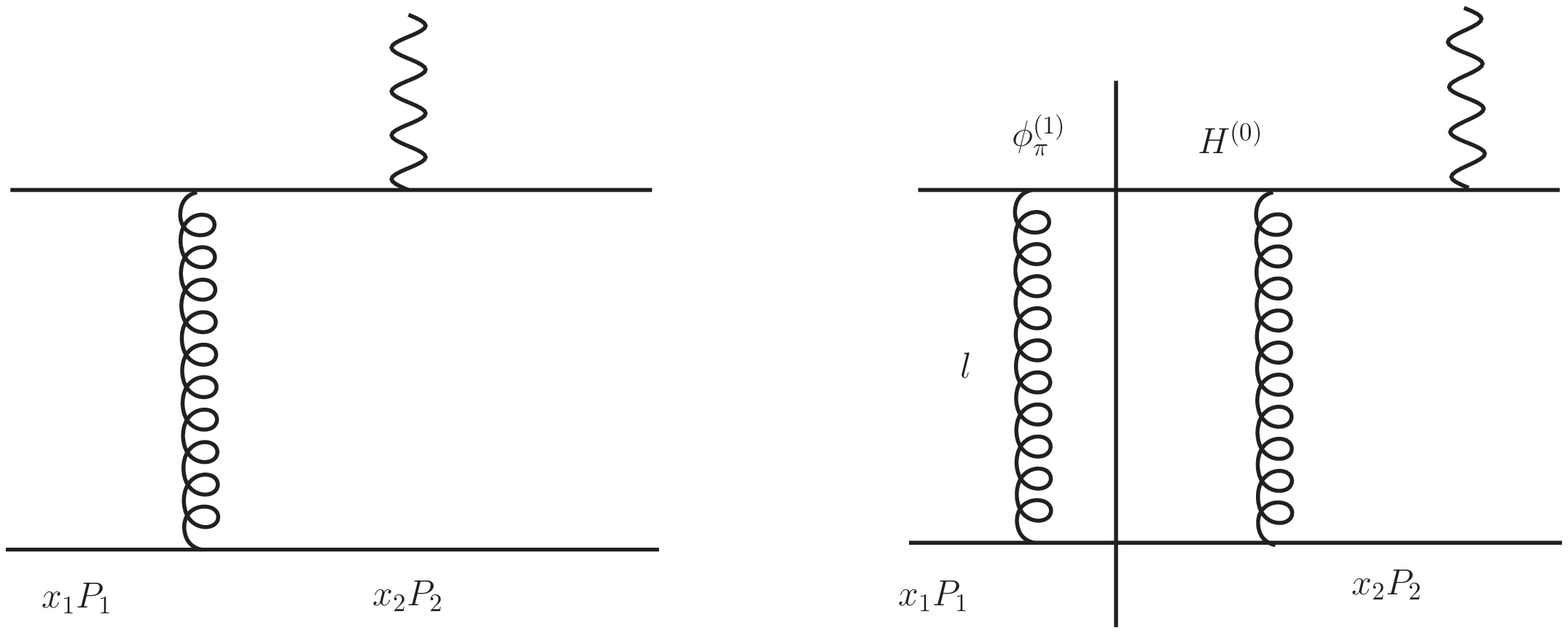}}
\begin{center}
\caption{(a) Lowest-order diagram for $F_\pi$. (b) Radiative
correction to (a).}
\end{center}
\label{fig:pionform}
\end{figure}
%%%%%%%%%%%%%%%%%%%%%%%%%%%
An example of next-to-leading-order diagrams is shown in Fig.~2(b).
The factorization of Fig.~2(b) is trivial: one performs the Fierz 
transformation to separate the fermion flows, so that the right-hand
side of the cut corresponds to the lowest-order hard kernel $H^{(0)}$. 
Since the loop momentum $l$ flows into the hard gluon, we have the gluon 
momentum $x_1P_1-x_2P_2+l$ and
\begin{eqnarray}
H^{(0)}\propto\frac{-1}{(x_1P_1-x_2P_2)^2+2x_1P_1^+l^-
-2x_2P_2^-l^++2l^+l^--l_T^2}\;.
\end{eqnarray}
Dropping $l^-$ and $l_T$ as a collinear approximation, the above
expression reduces to
\begin{eqnarray}
H^{(0)}(\xi_1,x_2)\propto\frac{1}{2x_1x_2P_1^+P_2^-+2x_2P_2^-l^+}
\equiv\frac{1}{\xi_1x_2Q^2}\;,
\end{eqnarray}
where $\xi_1=x_1+l^+/P_1^+$ is the parton momentum fraction modified by
the collinear gluon exchange. The left-hand side of the cut then
contributes to the $O(\alpha_s)$ distribution amplitude 
$\phi_\pi^{(1)}(\xi_1)$, which contains the integration over $l^-$ and 
$l_T$. Therefore, factorization to all orders gives a convolution only in
the longitudinal components of parton momentum,
\begin{eqnarray}
F_\pi=\int d\xi_1 d\xi_2\phi_\pi(\xi_1)H(\xi_1,\xi_2)
\phi_\pi(\xi_2)\;.
\end{eqnarray}

In the region with small parton momentum fractions, the hard scale
$x_1x_2Q^2$ is not large. In this case one may
drop only $l^-$, and keep $l_T$ in $H^{(0)}$. This weaker approximation
gives \cite{NL}
\begin{eqnarray}
H^{(0)}(\xi_1,x_2,l_T)\propto
\frac{1}{2(x_1+l^+/P_1^+)x_2P_1^+P_2^-+l_T^2}
\equiv\frac{1}{\xi_1x_2Q^2+l_T^2}\;,
\end{eqnarray}
which acquires a dependence on a transverse
momentum. We factorize the left-hand side of the cut in Fig.~2(b) into
the $O(\alpha_s)$ wave function $\phi_\pi^{(1)}(\xi_1,l_T)$, which
involves the integration over $l^-$. It is understood that the collinear 
gluon exchange not only modifies the momentum fraction, but introduces 
the transverse momentum dependence of the pion wave function.
Extending the above procedure to all orders, we derive the $k_T$
factorization,
\begin{eqnarray}
F_\pi=\int d\xi_1 d\xi_2 d^2k_{1T} d^2k_{2T}
\phi_\pi(\xi_1,k_{1T})H(\xi_1,\xi_2,k_{1T},k_{2T})
\phi_\pi(\xi_2,k_{2T})\;.
\end{eqnarray}
%which contains a convolution in both the longitudinal and transverse
%components of parton momenta. It has been shown that predictions derived
%from $k_T$ factorization theorem are gauge invariant \cite{NL}. 

\section{Ingredients of $k_T$ Factorization Approach  }

{\bf   Factorization in PQCD:}
The idea of pertubative QCD is as follows:
When heavy B-meson decays into two light mesons, the hard process is
dominant. Since two light mesons fly so fast with large momentum,
it is reasonable assumptions that the final-state interaction is not
important for charmless B-decays. Hard gluons are needed to boost
the resting spectator quark to get large momentum and finally 
to hadronize a fast moving final meson. 
So the dominant process is that one hard gluon
is exchanged between specator quark and other four quarks.

Let's start with the lowest-order diagram 
of $B \to K\pi$. The soft divergences in the $B \to \pi$ form factor
can be factorized into a light-cone B meson wave function,
and the collinear divergences can be absorbed into 
a pion distribution amplitude.
The finite pieces of them is absorbed into the hard part.
Then in the natural way we can factorize amplitude into two pieces:
$G \equiv H(Q,\mu) \otimes \Phi(m, \mu)$ where H stands for hard part
which is calculable with a perturbative way. $\Phi$ represents 
a product of wave functions which contains all the nonperturbative dynamics.

PQCD adopt the three scale factorization theorem \cite{Li:01}
based on the perturbative QCD formalism by Brodsky and Lepage \cite{BL},
and Botts and Sterman \cite{BS}, with the inclusion of the transverse
momentum components carried by partons inside meson.

We have three different scales: electroweak scale: $M_W$,
hard interaction scale: $t \sim O( \sqrt( \bar{\Lambda}m_b))$, 
and the factorization scale: $1/b$ where
$b$ is the conjugate variable of parton transverse momenta.
The dynamics below $1/b$ is completely non-perturbative and 
can be parameterized into meson wave funtions which are universal and
process independent. In our analysis
we use the results of light-cone distribution amplitudes (LCDAs)
by Ball \cite{PB:01,PB:02} with the light-cone sum rule calculation.

The ampltitude in PQCD is expressed as 
\begin{eqnarray}
\langle M_{1} M_{2}|C_{k}(t){\cal O}_{k}|B\rangle &=&\int [dx]\int
\left[ \frac{d^{2}\vec{b}}{ 4\pi }\right] \Phi
_{M_{1}}^{*}(x_{2},\vec{b} _{2})\,\Phi _{M_{2}}^{*}(x_{3},\vec{b}
_{3})\, C_{k}(t) \nonumber \\ 
&\otimes & H_{k}(\{x\},\{\vec{b}\},M_{B})\Phi
_{B}(x_{1},\vec{b}_{1}) \,\,{S_{t}( \{x\})} \,\,
{e^{-S\left( \{x\},\{\vec{b}\},M_{B}\right) }}
\end{eqnarray}
with the sudakov suppressed factor:
$S=S_{B}(x_{1}P^{+}_{1},b_{1})+S_{M_{1}}(x_{2}P^{-}_{2},b_{2})
+S_{M_{1}}((1-x_{2})P^{-}_{2},b_{2})+...$
and the threshold resummation factor $S_t(x)$.
Here $C(t)$ are Wilson coefficients, $\Phi(x)$ are meson LCDAs
and variable $t$ is the factorized scale in hard part.
%%%%%  Figure 2 %%%%%%%%%%%%%%%%%%%%%%%
%%%%%%%%%%%%%%%%%%%%% sudakov resummation  %%%%%%%%%%%%%%%%%%%%%%%%
\vspace{3mm}
\begin{center}
\vspace{-30pt} \hfill \\
\begin{picture}
(70,0)(70,25)
\ArrowLine(80,10)(30,10)
\ArrowLine(130,10)(80,10)
\ArrowLine(30,-30)(80,-30)
\ArrowLine(80,-30)(130,-30)
\Gluon(80,10)(80,-30){4}{4}
\GOval(30,-10)(20,10)(0){0.5}
%\Text(80,-60)[]{$(c)$}
\end{picture}\hspace{20mm}
\begin{picture}
(70,0)(70,25)
\ArrowLine(80,10)(30,10)
\ArrowLine(130,10)(80,10)
\ArrowLine(30,-30)(80,-30)
\ArrowLine(80,-30)(130,-30)
\GlueArc(80,-30)(15,0,180){4}{4}
\GOval(30,-10)(20,10)(0){0.5}
\end{picture} 
\end{center}

\vskip1.5cm
\begin{figure}[h]
\caption{The diagrams generate double logarithm corrections 
for the sudakov resummation.}
\end{figure}
%%%%%%%%%%%%%%%%%%%%%%%%%%%%%%%%%%%%%%%%%%%%%%%%%%%%%%%%%%%%%%%%%

\vspace{3mm}
{\bf Sudakov Suppression Effects:} 
When we include $k_{\perp}$,
the double logarithms $\ln^2(Pb)$ are generated 
from the overlap of collinear and soft divergence in radiative corrections
to meson wave functions(See figure 3),
where P is the dominant light-cone component of a meson momentum. 
The resummation of these double logarithms leads to a Sudakov form factor
$exp[-s(P,b)]$ in Eq.(8), which suppresses the long distance contributions
in the large $b$ region, and vanishes as $b > 1/\Lambda_{QCD}$.
%%%%%%%%%%%%%%%%%%%%%%%%%%%%%%%%%%%%%%%%%%%%%
\begin{figure}[ht]
\centerline{\epsfxsize2.5 in \epsffile{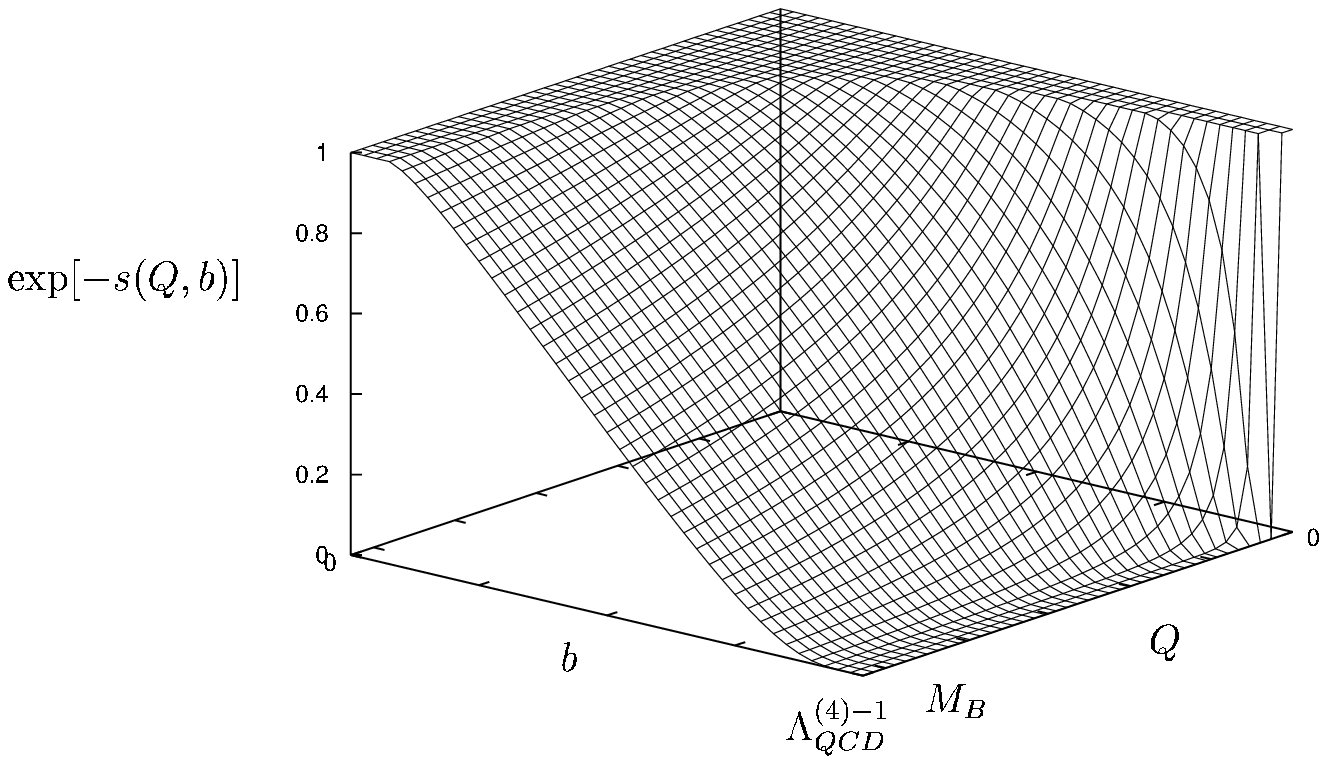}\hspace{1.0cm}
\epsfxsize2.5 in \epsffile{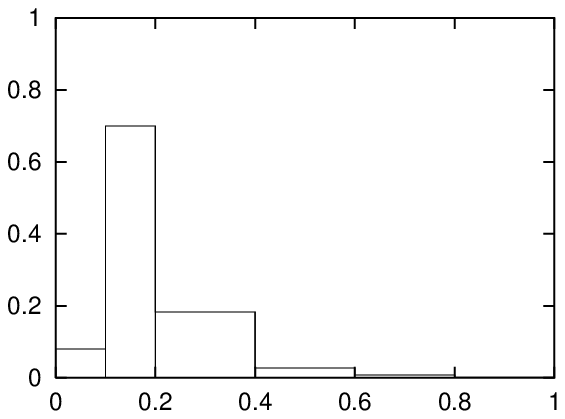} }
%  \begin{picture}(0,0)(0,0)
%   \put(-50,-5){${\alpha_s(t)}/{\pi}$}
%   \put(-185,80){\rotatebox{90}{Fraction}}
%  \end{picture}
\caption{(a)Sudakov suppression factor (b)Fractional contribution to
  the $B \to \pi$ transition form factor $F^{B\pi}$ as a function of
$\alpha_s(t)/\pi$.} 
\end{figure}
%%%%%%%%%%%%%%%%%%%%%%%%%%%%%%%%%%%%%%%%%%%%%%
This suppression renders $k_{\perp}^2$ flowing into the hard amplitudes
of order
\begin{eqnarray}
<k_{\perp}^2> \sim O(\bar\Lambda M_B)\;.
\end{eqnarray}
The off-shellness of internal particles then remain of
$O(\bar\Lambda M_B)$ even in the end-point region, and the singularities
are removed. This mechanism is so-called Sudakov suppression(See figure 4-a).

Du {\it et al.} have studied the Sudakov effects in the evaluation
of nonfactorizable amplitudes \cite{Du}. If equating these amplitudes
with Sudakov suppression included to the parametrization in QCDF, it was
observed that the corresponding cutoffs are located in the reasonable
range proposed by Beneke {\it et al.} \cite{BBNS:00}.
Sachrajda {\it et al.} have expressed an opposite opinion on the effect
of Sudakov suppression in \cite{GS}. However, their conclusion was drawn
based on a very sharp $B$ meson wave function, which is not favored by
experimental data.

Here I would like to comment on the negative opinions on the large
$k_{\perp}^2 \sim O(\bar{\Lambda}M_B)$.
It is easy to understand the increase of $k_{\perp}^2$ from $O(\bar\Lambda^2)$,
carried by the valence quarks which just come out of the initial meson
wave functions, to $O(\bar\Lambda M_B)$, carried by the quarks which are
involved in the hard weak decays. Consider the simple deeply inelastic
scattering of a hadron. The transverse momentum $k_{\perp}$ carried by a
parton, which just come out of the hadron distribution function, is
initially small. After infinite many gluon radiations, $k_{\perp}$ becomes of
$O(Q)$, when the parton is scattered by the highly virtual photon,
where $Q$ is the large momentum transfer from the photon. The evolution
of the hadron distribution function from the low scale to $Q$ is described
by the Dokshitzer-Gribov-Lipatov-Altarelli-Parisi (DGLAP) equation
\cite{GL,AP}. The mechanism of the DGLAP evolution in DIS is similar to that
of the Sudakov evolution in exclusive $B$ meson decays. The difference
is only that the former is the consequence of the single-logarithm
resummation, while the latter is the consequence of the double-logarithm
resummation.

By including Sudakov effects, all contributions of the $B \to \pi$
form factor comes from the region with $\alpha_s/\pi < 0.3$ \cite{KLS:02}
as shown in Figure~4(b). It indicate that our PQCD results are well within
the perturbative region. 
%%%%%  Figure 3 %%%%%%%%%%%%%%%%%%%%%%%
%%%%%%%%%%%%%%%%%%%%% threshold resummation  %%%%%%%%%%%%%%%%%%%%%%%%
\begin{center}
\vspace{-30pt} \hfill \\
\begin{picture}
(70,0)(70,25)
\ArrowLine(80,10)(30,10)
\ArrowLine(130,10)(80,10)
\ArrowLine(30,-30)(80,-30)
\ArrowLine(80,-30)(130,-30)
\Gluon(50,10)(50,-30){3}{6}
\GBoxc(80,10)(7,7){0}
\GlueArc(80,10)(15,180,360){3}{6}
%\Text(80,-60)[]{$(c)$}
\end{picture}\hspace{25mm}
\begin{picture}
(70,0)(70,25)
\ArrowLine(80,10)(30,10)
\ArrowLine(130,10)(80,10)
\ArrowLine(30,-30)(80,-30)
\ArrowLine(80,-30)(130,-30)
\Gluon(110,10)(110,-30){3}{6}
\GBoxc(80,10)(7,7){0}
\GlueArc(80,10)(15,180,360){3}{6}
%\Text(80,-60)[]{$(d)$}
\end{picture} 
\end{center}
\vskip 2.0cm
\begin{figure}[ht]
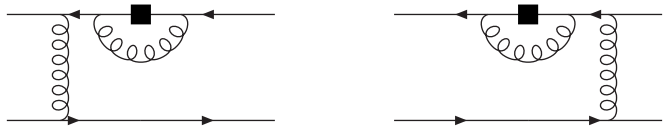

\caption{The diagrams generate double logarithm corrections 
for the threshold resummation.}
\end{figure}
%%%%%%%%%%%%%%%%%%%%%%%%%%%%%%%%%%%%%%%%%%%%%%%%%%%%%%%%%%%%%%%%%

\vspace{3mm}
{\bf Threshold Resummation:} 
The other double logarithm is $\alpha_s \ln^2(1/x)$ from the end point region
of the momentum fraction $x$ \cite{Li:02}. This double logarithm  is generated
by the corrections of the hard part in Figure 5.
This double logarithm can be factored out of the hard amplitude
systematically, and its resummation introduces a Sudakov factor 
$S_t(x)=1.78 [x(1-x)]^c$ with $c=0.3$ into PQCD factorization formula.
The Sudakov factor from threshold resummation 
is universal, independent of flavors of internal quarks, the twists and topologies 
of hard amplitudes, and the decay modes.

Threshold resummation\cite{Li:02} and $k_{\perp}$ resummation 
\cite{CS,BS,StLi} 
arise from different
subprocesses in PQCD factorization and suppresses the
end-point contributions, making PQCD evaluation of exclusive $B$ meson
decays reliable. Without these resummation effects, the PQCD predictions
for the $B\to K$ form factors are infrared divergent. If including only
$k_{\perp}$ resummation, the PQCD predictions are finite. However, the
two-parton twist-3 contributions are still huge, so that the $B\to K$
form factors have an unreasonably large value $F^{BK}\sim 0.57$ at maximal
recoil. The reason is that the double logarithms $\alpha_s\ln^2 x$ have
not been organized. If including both resummations, we obtain the
reasonable result $F^{BK}\sim 0.35$ as shown in Figure~6.. 
These studies indicate the importance
of resummations in PQCD analyses of $B$ meson decays. 
In conclusion, if the PQCD analysis of the heavy-to-light form factors is performed
self-consistently, there exist no end-point singularities, and both
twist-2 and twist-3 contributions are well-behaved.

%%%%%%%%%%%%%%%%%%%%%%%%%%%%%%%%%%%%%%%
\begin{table}[hb]
\begin{tabular}{|c|cc|c|} \hline 
Amplitudes & twist-2 contribution & 
Twist-3 contribution & Total \\
\hline 
$Re(f_{\pi} F^T)$ & \hspace{0.5cm}$3.44 \cdot 10^{-2}$ & 
\hspace{0.5cm}$5.00 \cdot 10^{-2}$
 & $8.44 \cdot 10^{-2}$   \\
$Im(f_{\pi} F^T)$ & $-$  & $-$ &  $-$ \\ 
\hline
$Re(f_{\pi} F^P)$ &  \hspace{0.5cm}-$1.26 \cdot 10^{-3}$ & 
\hspace{0.5cm}-$4.76 \cdot 10^{-3}$ 
& -$6.02 \cdot 10^{-3}$ \\
$Im(f_{\pi} F^P)$ & $-$ & $-$ & $-$ \\
\hline
$Re(f_{B} F_a^P)$ & \hspace{0.5cm}$2.52 \cdot 10^{-6}$ & 
\hspace{0.5cm}-$3.30 \cdot 10^{-4}$
& -$3.33 \cdot 10^{-4}$  \\
$Im(f_{B} F_a^P)$ & \hspace{0.5cm}$8.72 \cdot 10^{-7}$ & 
\hspace{0.5cm}$3.81 \cdot 10^{-3}$ 
& $3.81 \cdot 10^{-3}$ \\
\hline 
$Re(M^T)$ & \hspace{0.5cm}$7.26 \cdot 10^{-4}$ & 
\hspace{0.5cm}-$1.39 \cdot 10^{-6}$ 
& -$7.25 \cdot 10^{-4}$ \\
$Im(M^T)$ & \hspace{0.5cm}-$1.62 \cdot 10^{-3}$ & 
\hspace{0.5cm}-$2.91 \cdot 10^{-4}$
& $1.33 \cdot 10^{-3}$  \\
\hline
$Re(M^P)$ & \hspace{0.5cm}-$1.67 \cdot 10^{-5}$ & 
\hspace{0.5cm}-$1.47 \cdot 10^{-7}$
& $1.66 \cdot 10^{-5}$  \\
$Im(M^P)$ & \hspace{0.5cm}-$3.52 \cdot 10^{-5}$ & 
\hspace{0.5cm} $6.56 \cdot 10^{-6}$
& -$2.87 \cdot 10^{-5}$ \\
\hline
$Re(M_a^P)$ & \hspace{0.5cm}-$7.37 \cdot 10^{-5}$ & 
\hspace{0.5cm}$2.50 \cdot 10^{-6}$
& -$7.12 \cdot 10^{-5}$ \\
$Im(M_a^P)$ & \hspace{0.5cm}-$3.13 \cdot 10^{-5}$ & 
\hspace{0.5cm}-$2.04 \cdot 10^{-5}$ 
& -$5.17 \cdot 10^{-5}$ \\ \hline
\end{tabular}
\label{TABLE11.1}
\caption{Amplitudes for the $B_d^{0} \to \pi^{+} \pi^{-}$ decay 
where $F$ ($M$) denotes factorizable (nonfactorizable) 
contributions, $P$ ($T$) denotes the penguin (tree) contributions,
and $a$ denotes the annihilation contributions. Here we adopted 
$\phi_3=80^0$, $R_b=0.38$, $m_0^{\pi}=1.4 \,GeV$ and
$\omega_B=0.40 \, GeV$.  }
\end{table} 
%%%%%%%%%%%%%%%%%%%%%%%%%%%%%%%%%%%%%%%
%%%%%%%%%%%%%%%%%%%%%%%%%%%%%%%%%%%%%%%%%%%%%
\begin{figure*}[ht]
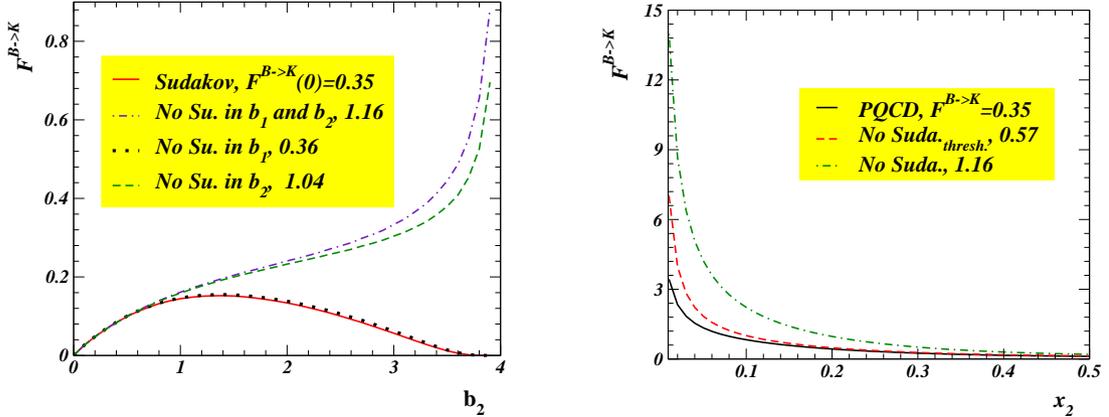

\begin{center}
\includegraphics[angle=0,width=0.4\textwidth]{fkb.eps}
\hspace{10mm}
\includegraphics[angle=0,width=0.4\textwidth]{fkx.eps}
%\centerline{\epsfxsize2.5 in \epsffile{fkb.eps}\hspace{1.0cm}
%\epsfxsize2.5 in \epsffile{fkx.eps} }
\caption{Sudakov suppression and threshold resummation effects in 
$B \to K$ transition form factor}
\end{center} 
\end{figure*}
%%%%%%%%%%%%%%%%%%%%%%%%%%%%%%%%%%%%%%%%%%%%%%
%%%%%%%%%%%%%%%%%%%%%%%%%%%%%%%%%%%%%%%%%%%%
\begin{figure}[ht]
\epsfxsize=12.cm
\centerline{ \epsfbox{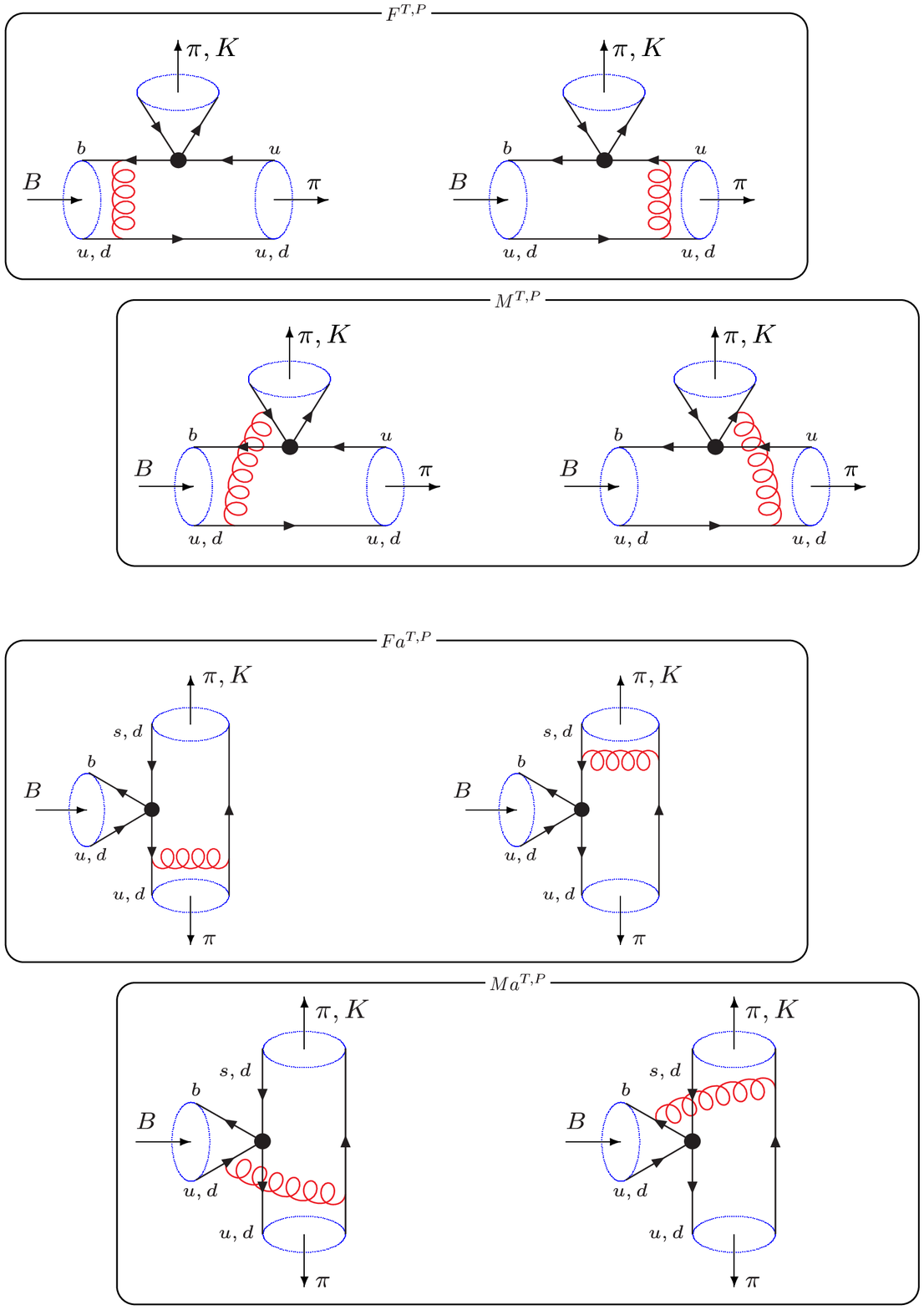}}
\begin{center}
\caption{Feynman diagrams for $B \to \pi\pi$ and $K\pi$.}
\end{center}
\label{fig:feynman}
\end{figure}
%%%%%%%%%%%%%%%%%%%%%%%%%%%%%%%%%%%%%%%%%%%%%%
\vspace{5mm}
{\bf Power Counting Rule in PQCD:} 
The power behaviors of various topologies of diagrams for two-body
nonleptonic $B$ meson decays with the Sudakov effects taken into account
has been discussed in details in \cite{CKL:a}. The relative importance is
summarized below:
\begin{eqnarray}
{\rm emission} : {\rm annihilation} : {\rm nonfactorizable} 
=1 : \frac{2m_0}{M_B} : \frac{\bar\Lambda}{M_B}\;,
\label{eq1}
\end{eqnarray}
with $m_0$ being the chiral symmetry breaking scale. The scale $m_0$
appears because the annihilation contributions are dominated by those
from the $(V-A)(V+A)$ penguin operators, which survive under helicity
suppression. In the heavy quark limit the annihilation and
nonfactorizable amplitudes are indeed power-suppressed compared to the
factorizable emission ones. Therefore, the PQCD formalism for two-body
charmless nonleptonic $B$ meson decays coincides with the factorization
approach as $M_B\to\infty$. However, for the physical value $M_B\sim 5$
GeV, the annihilation contributions are essential.
In Table 1 and 2 we can easily check the relative size of the different topology
in Eq.(\ref{eq1}) by the peguin contribution for W-emission
($f_{\pi}F^{P}$), annihilation($f_BF^{P}_a$) and
non-factorizable($M^P$) contributions as shown in Figure 7. 
Specially we show the relative
size of the different twisted light-cone-distribution-amplitudes (LCDAs)
for each topology. We have more sizable twist-3 contributions in
the factorizable diagram.

Note that all the above topologies are of the same order in $\alpha_s$
in PQCD. The nonfactorizable amplitudes are down by a power of $1/m_b$,
because of the cancellation between a pair of nonfactorizable diagrams,
though each of them is of the same power as the factorizable one. I
emphasize that it is more appropriate to include the nonfactorizable
contributions in a complete formalism. The
factorizable internal-$W$ emisson contributions are strongly suppressed
by the vanishing Wilson coefficient $a_2$ in the $B\to J/\psi K^{(*)}$
decays \cite{YL}, so that nonfactorizable contributions become
dominant\cite{charmonium}.
In the $B\to D\pi$ decays, there is no soft cancellation between a pair
of nonfactorizable diagrams, and nonfactorizable contributions are
significant \cite{YL}.

In QCDF the factorizable and nonfactorizable amplitudes are of the same
power in $1/m_b$, but the latter is of next-to-leading order in
$\alpha_s$ compared to the former. Hence, QCDF approaches FA in the
heavy quark limit in the sense of $\alpha_s\to 0$. Briefly speaking,
QCDF and PQCD have different counting rules both in $\alpha_s$ and in
$1/m_b$. The former approaches FA logarithmically
($\alpha_s\propto 1/\ln m_b \to 0$), while the latter does linearly
($1/m_b\to 0$).

%%%%%%%%%%%%%%%%%%%%%%%%%%%%%%%%%%%%%%%%%%%%%%%%%%%%%%%%%%%
\begin{table}[hbt]
\begin{tabular}{|c|cc|c|} \hline 
Amplitudes & Left-handed gluon exchange & 
Right-handed gluon exchange & Total \\
\hline 
$Re(f_{K} F^T)$ & \hspace{0.5cm}$7.07 \cdot 10^{-2}$ & 
\hspace{0.5cm}$3.16 \cdot 10^{-2}$
 & $1.02 \cdot 10^{-1}$   \\
$Im(f_{K} F^T)$ & $-$  & $-$ &  $-$ \\ 
\hline
$Re(f_{K} F^P)$ &  \hspace{0.5cm}-$5.52 \cdot 10^{-3}$ & 
\hspace{0.5cm}-$2.44 \cdot 10^{-3}$ 
& -$7.96 \cdot 10^{-3}$ \\
$Im(f_{K} F^P)$ & $-$ & $-$ & $-$ \\
\hline
$Re(f_{B} F_a^P)$ & \hspace{0.5cm}$4.13 \cdot 10^{-4}$ & 
\hspace{0.5cm}-$6.51 \cdot 10^{-4}$
& -$2.38 \cdot 10^{-4}$  \\
$Im(f_{B} F_a^P)$ & \hspace{0.5cm}$2.73 \cdot 10^{-3}$ & 
\hspace{0.5cm}$1.68 \cdot 10^{-3}$ 
& $4.41 \cdot 10^{-3}$ \\
\hline 
$Re(M^T)$ & \hspace{0.5cm}$7.06 \cdot 10^{-3}$ & 
\hspace{0.5cm}-$7.17 \cdot 10^{-3}$ 
& -$1.11 \cdot 10^{-4}$ \\
$Im(M^T)$ & \hspace{0.5cm}-$1.10 \cdot 10^{-2}$ & 
\hspace{0.5cm}$1.35 \cdot 10^{-2}$
& $2.59 \cdot 10^{-3}$  \\
\hline
$Re(M^P)$ & \hspace{0.5cm}-$3.05 \cdot 10^{-4}$ & 
\hspace{0.5cm}$3.07 \cdot 10^{-4}$
& $2.17 \cdot 10^{-6}$  \\
$Im(M^P)$ & \hspace{0.5cm}$4.50 \cdot 10^{-4}$ & 
\hspace{0.5cm}-$5.29 \cdot 10^{-4}$
& -$7.92 \cdot 10^{-5}$ \\
\hline
$Re(M_a^P)$ & \hspace{0.5cm}$2.03 \cdot 10^{-5}$ & 
\hspace{0.5cm}-$1.37 \cdot 10^{-4}$
& -$1.16 \cdot 10^{-4}$ \\
$Im(M_a^P)$ & \hspace{0.5cm}-$1.45 \cdot 10^{-5}$ & 
\hspace{0.5cm}-$1.27 \cdot 10^{-4}$ 
& -$1.42 \cdot 10^{-4}$ \\ \hline
\end{tabular}
\label{TABLE11.3}
\caption{Amplitudes for the $B_d^{0} \to K^{+} \pi^{-}$ decay 
where $F$ ($M$) denotes factorizable (nonfactorizable) 
contributions, $P$ ($T$) denotes the penguin (tree) contributions,
and $a$ denotes the annihilation contributions. Here we adopted 
$\phi_3=80^0$, $R_b=0.38$.  }
\end{table} 
%%%%%%%%%%%%%%%%%%%%%%%%%%%%%%%%%%%%%%%%%%%%%%%%%%%%%%%%%%%%%%%%%%%%%%%%%

\section{Important Theoretical Issues}
{\bf End Point Singularity and Form Factors:} 
If calculating the $B\to\pi$ form factor $F^{B\pi}$ at large recoil using
the Brodsky-Lepage formalism \cite{BL,BSH}, a difficulty immediately
occurs. The lowest-order diagram for the hard amplitude is proportional to 
$1/(x_1 x_3^2)$, $x_1$ being the momentum fraction associated with the
spectator quark on the $B$ meson side. If the pion distribution amplitude
vanishes like $x_3$ as $x_3\to 0$ (in the leading-twist, {\it i.e.},
twist-2 case), $F^{B\pi}$ is logarithmically divergent. If the pion
distribution amplitude is a constant as $x_3\to 0$ (in the
next-to-leading-twist, {\it i.e.}, twist-3 case), $F^{B\pi}$ even becomes
linearly divergent. These end-point singularities have also appeared in
the evaluation of the nonfactorizable and annihilation amplitudes in QCDF
mentioned above.

When we include small parton transverse momenta $k_{\perp}$, we have
\begin{equation}
{1 \over x_1\,\, x_3^2 M_B^4} \hspace{10mm} \rightarrow
\hspace{10mm} {1 \over (x_3\, M_B^2 + k_{3\perp}^2) \,\,
[x_1x_3\, M_B^2 + (k_{1\perp} - k_{3\perp})^2]}
\label{eq:4} 
\end{equation}
and the end-point singularity is smeared out.

In PQCD, we can calculate analytically space-like form factors for $B \to P,V$
transition and
also time-like form factors for the annihilation process \cite{CKL:a,Kurimoto}.

\vspace{3mm}
{\bf Strong Phases:} 
While stong phases in FA and QCDF 
come from the Bander-Silverman-Soni (BSS) mechanism\cite{BSS}
and from the final state interaction (FSI), the dominant strong phase in PQCD
come from the factorizable annihilation
diagram\cite{KLS:01,KLS:02,KLS:03}
(See Figure 8). In fact,
the two sources of strong phases in the FA
and QCDF approaches are strongly suppressed by the charm mass
threshold and by the end-point behavior of meson wave functions.
So the strong phase in QCDF is almost zero without soft-annihilation
contributions.

%%%%%%%%%%%%%%%%%%%%%%%%%%
\vskip0.5cm
\begin{figure}[htbp]
\epsfxsize=10cm
\centerline{\epsfbox{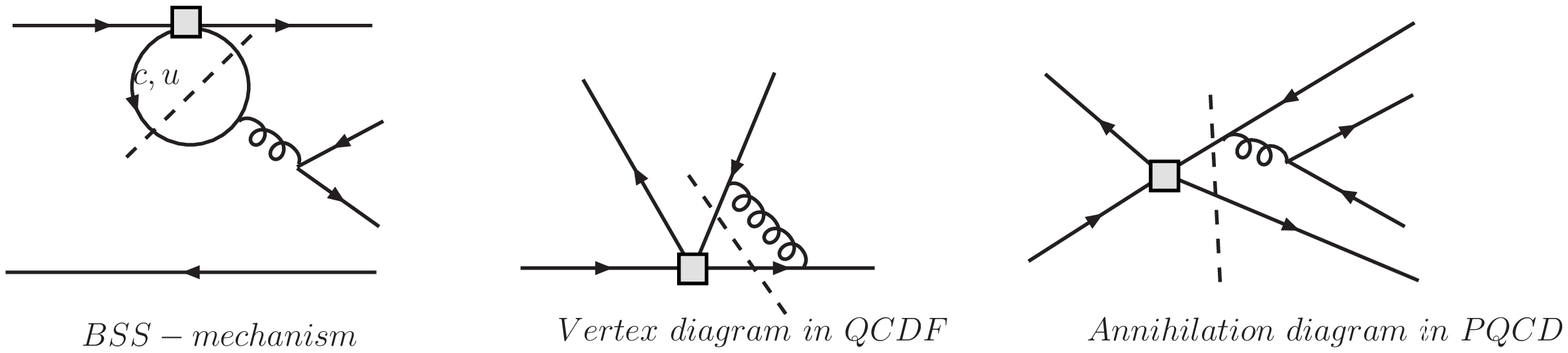}}
\begin{center}
\caption{Different sourses of strong phase: (a)BSS mechanism, 
(b) Final State Interaction, and 
(c) Factorizable annihilation. }
\end{center}
\label{fig:strongphase}
\end{figure}
%%%%%%%%%%%%%%%%%%%%%%%%%%%

{\bf Dynamical Penguin Enhancement vs Chiral Enhancement:} 
As explained before, the hard scale is about 1.5 GeV.
Since the RG evolution of the Wilson coefficients $C_{4,6}(t)$ increase
drastically as $t < M_B/2$, while that of $C_{1,2}(t)$ remain almost
constant as shown in Figure~9, 
we can get a large enhancement effects from both wilson
coefficents and matrix elements in PQCD. 
 
In general the amplitude can be expressed as
\begin{equation}
Amp \sim [a_{1,2} \,\, \pm \,\, a_4 \,\,
\pm \,\, m_0^{P,V}(\mu) a_6] \,\, \cdot \,\, <K\pi|O|B>
\label{eq:2}
\end{equation}
with the chiral factors $m_0^P(\mu)=m_P^2/[m_1(\mu)+m_2(\mu)]$ for
pseudoscalr meson 
and $m_0^{V}= m_V$ for vector mesons.
To accommodate the $B\to K\pi$ data in the factorization and
QCD-factorization approaches, one relies on the chiral enhancement by
increasing the mass $m_0$ to as large values about 3 GeV at $\mu=m_b$ scale.
So two methods accomodate large branching ratios of $B \to K\pi$ and
it is difficult for us to distinguish two different methods in $B \to
PP$ decays. However we can do it in $B \to PV$ because there is no
chiral factor in LCDAs of the vector meson. 
%%%%%%%%%%%%%%%%%%%%%%%%%%
\begin{figure}
\epsfxsize=8cm
\centerline{\epsfbox{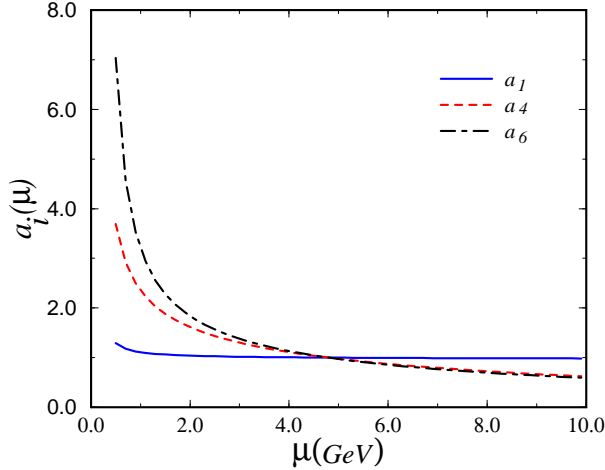}}
\begin{center}
\caption{Dynamical enhancement of Wilson coefficents $a_i$ (i=1,4,6).}
\end{center}
\label{fig:wilsonai}
\end{figure}
%%%%%%%%%%%%%%%%%%%%%%%%%%%
We can test whether dynamical enhancement 
or chiral enhancement is responsible
for the large $B \to K\pi$ branching ratios 
by measuring the $B \to \phi K$ modes.
In these modes penguin contributions dominate, 
such that their branching ratios are
insensitive to the variation of the unitarity angle $\phi_3$.
According to recent
works by Cheng {\it at al.} \cite{CK}, 
the branching ratio of $B \to \phi K$ is $(2-7)
\times 10^{-6}$ including $30\%$ annihilation contributions in the
QCD-factorization approach (QCDF). 
However PQCD predicts $10 \times 10^{-6}$ \cite{CKL:a,Mishima}.   
For $B \to \phi K^{*}$ decays, QCDF gets about $9 \times 10^{-6}$\cite{HYC},
but PQCD have $15 \times 10^{-6}$\cite{CKL:b}.
Because of these small branching ratios for $B\to PV$ and $VV$ decays
in the QCD-factorization approach, they can not globally fit the
experimental data for $B\to PP,VP$ and $VV$ modes simultaneously
with same sets of free parameters $(\rho_H,\phi_H)$ and $(\rho_A,\phi_A)$
\cite{zhu}.

\vspace{5mm}
{\bf Fat Imaginary Penguin in Annihilation:} 
There is a falklore that the annihilation contribution is negligible
compared to W-emission one. In this reason the annihilation contribution
was not included in the general factorization approach and the first
paper on QCD-factorization by Beneke et al. \cite{BBNS:99}.
In fact there is a suppression effect for the operators with structure
$(V-A)(V-A)$ because of a mechanism similar to the helicity
suppression for $\pi \to \mu \nu_{\mu}$. However annihilation from 
the operators $O_{5,6,7,8}$ with the structure $(S-P)(S+P)$ via Fiertz
transformation survive under the helicity suppression and can get
large imaginary value. The real part of factorized annihilation contribution
becomes small because there is a cancellation between left-handed
gluon exchanged one and right-handed gluon exchanged one as shown in
Table 2. This mostly pure imaginary value of annihilation is a main
source of large CP asymmetry in $B \to \pi^{+}\pi^{-}$ and $K^{+}\pi^{-}$.
In Table 6 we summarize the CP asymmetry in 
$B \to K(\pi)\pi$ decays.
%%%%%%%%%%%%%%%%%%%%%%%%%%%%%%%%%%%%%%%%%%%%%%%%%%%%%%%%%%%%%%%%%%%%%%%%

%%%%%%%%%%%%%%%%%%%%%%%%%%%%%%%%%%%%%%%%%%%%%%%%%%%%%%%%%%%%%%%%%%%%%%%%%
\section{Numerical Results}
{\bf Branching ratios in Charmless B-decays:} 
The PQCD approach allows us to calculate 
the amplitudes for charmless B-meson decays
in terms of ligh-cone distribution amplitudes upto twist-3. 
We focus on decays
whose branching ratios have already been measured. 
We take allowed ranges of shape parameter for the B-meson wave funtion as 
$\omega_B = 0.36-0.44$ which accomodate to reasonable form factors, 
$F^{B\pi}(0)=0.27-0.33$ and $F^{BK}(0)=0.31-0.40$. 
We use values of chiral factor
with $m_0^{\pi}=1.3 GeV$ and $m_0^{K}=1.7 GeV$.
Finally we obtain branching ratios for $B\to K(\pi)\pi$ 
\cite{KLS,LUY}, 
$K\phi$ \cite{CKL:a,Mishima} $K^{*}\phi$\cite{CKL:b} and 
$K^{*}\pi$\cite{Keum02},
which is well agreed with present experimental data in Table 3-5.
%%%%%%%%%%%%%%%%%%%%%%% Branching ratios (1) %%%%%%%%%%%%%%%%%%%%
\small
\begin{table}[t]
\caption{Branching ratios of $B \to \pi\pi, K\pi$and $KK$ decays 
with $\phi_3=80^0$, $R_b=\sqrt{\rho^2+\eta^2}=0.38$. 
Here we adopted $m_0^{\pi}=1.3$ GeV,
$m_0^{K}=1.7$ GeV and $0.36<\omega_B<0.44$.
Unit is $10^{-6}$.} 
\label{TABLE1}
%\begin{center}
\begin{tabular} {|c|ccc|c|c|} \hline 
Modes & CLEO & BELLE & BABAR & ~~~World Av.~~~ & ~~~PQCD~~~  \\
\hline  
$\pi^{+}\pi^{-}$ & $4.5^{+1.4+0.5}_{-1.2-0.4}$ &
 $4.4\pm 0.6 \pm 0.3$ &
 $4.7\pm 0.6 \pm 0.2$ &  
$4.55 \pm 0.44$ &
$5.93-10.99$  \\
$\pi^{+}\pi^{0}$ & $4.5^{+1.8+0.6}_{-1.6-0.7}$ & 
 $5.3 \pm 1.3 \pm0.5$ &
 $5.8\pm 0.6 \pm 0.4$ & $5.20\pm0.79$ & 
$2.72-4.79$    \\ 
$\pi^{0}\pi^{0}$ & $<4.4$ & 
 $2.32^{+0.44+0.22}_{-0.48-0.18}$ &  $1.7\pm 0.32 \pm 0.10$ & $2.01\pm 0.43$ &
  $0.1-0.65$    \\ 
\hline
$K^{\pm}\pi^{\mp}$ &  
 $18.0^{+2.3+1.2}_{-2.1-0.9}$ &
 $18.5 \pm1.0 \pm0.7$ &  
 $17.9\pm 0.9 \pm 0.7$ &
 $18.2 \pm 0.8$ &  
 $12.67-19.30$    \\ 
$K^{0}\pi^{\mp}$ & 
 $18.8^{+3.7+2.1}_{-3.3-1.8}$ &
 $22.0 \pm1.9 \pm1.1$  &   
 $26.0\pm 1.3 \pm 1.0$ &
 $22.3 \pm 1.4$ &  
 $14.43-26.26$    \\ 
$K^{\pm}\pi^{0}$ &
 $12.9^{+2.4+1.2}_{-2.2-1.1}$ &
 $12.8 \pm1.4^{+1.4}_{-1.0}$ &  
 $12.0\pm 0.7\pm 0.6$ &
 $12.6 \pm 1.1$ &  
  $7.87-14.21$    \\
$K^{0}\pi^{0}$ &
 $12.8^{+4.0+1.7}_{-3.3-1.4}$ &
 $12.6 \pm2.4 \pm1.4$ &  
 $11.4 \pm 0.9 \pm 0.6$ &
 $12.3 \pm 1.7$ & 
 $4.46-8.06$    \\ 
\hline 
$K^{\pm}K^{\mp}$ &
 $<0.8$ &
 $<0.7$ &  
 $<0.6$ &
 $<0.6$ & 
 $0.06$    \\ 
$K^{\pm}\bar{K}^{0}$ &
 $<3.3$ &
 $<3.4$ &  
 $1.45\pm 0.50 \pm 0.11$ &
 $1.45\pm 0.50 \pm 0.11$ & 
 $1.4$    \\ 
$K^{0}\bar{K}^{0}$ &
 $<3.3$ &
 $<3.2$ &  
 $1.19 \pm 0.38 \pm 0.13$ &
 $1.19 \pm 0.38 \pm 0.13$ & 
 $1.4$    \\ 
\hline
\end{tabular}
%\end{center}
\end{table} 
%%%%%%%%%%%%%%%%%%%%%%%%%%%%%%%%%%%%%%%%%%%%%%%%%%%%%%%%%%%%%%%%%%
%%%%%%%%%%%%%%%%% Branching ratio (2)  %%%%%%%%%%%%%%%%%%%%%%%%%%%
\begin{table}[ht]
\caption{Branching ratios of $B \to \phi K^{(*)}$and $K^{*}\pi$ decays 
with $\phi_3=80^0$, $R_b=\sqrt{\rho^2+\eta^2}=0.38$. 
Here we adopted $m_0^{\pi}=1.3$ GeV
and $m_0^{K}=1.7$ GeV.
Unit is $10^{-6}$.} 
\label{TABLE2}
%\begin{center}
\begin{tabular}{|c|ccc|c|c|} \hline
Modes & CLEO & BELLE & BABAR 
& ~~ World Av.~~ &~~~PQCD~~~   \\
\hline  
$\phi K^{\pm}$ & 
 $5.5^{+2.1}_{-1.8}\pm 0.6$ &
 $9.4 \pm 1.1 \pm 0.7$ &  
 $10.0^{+0.9}_{-0.8}\pm 0.5$ &   
 $9.3 \pm 0.8$ &
 $8.1-14.1$  \\
$\phi K^{0}$ & 
 $ 5.4^{+3.7}_{-2.7} \pm 0.7 $ &
 $9.0 \pm 2.2 \pm 0.7$ &  
 $7.6^{+1.3}_{-1.2}\pm 0.5 $ &
 $7.7 \pm 1.1$  &
 $7.6-13.3$    \\ 
\hline
$\phi K^{*\pm}$ & 
 $10.6^{+6.4+1.8}_{-4.9-1.6}$ &  
 $6.7^{2.1+0.7}_{-1.9-1.0} $ & 
 $12.1^{+2.1}_{1.9} \pm 1.1$  &
 $9.4 \pm 1.6$ &
 $12.6-21.2$ \\
$\phi K^{*0}$ & 
 $11.5^{+4.5+1.8}_{-3.7-1.7} $ &
 $10.0^{+1.6+0.7}_{-1.5-0.8} $ &  
 $11.1^{+1.3}_{-1.2}\pm 0.8 $ &
 $10.7 \pm 1.1$    &
 $11.5-19.8$    \\ 
\hline
$K^{*0} \pi^{\pm}$ & 
 $7.6^{+3.5}_{-3.0} \pm 1.6$ &
 $19.4^{+4.2+4.1}_{-3.9-7.1}$ &  
 $15.5 \pm 3.4 \pm 1.8$ &
 $12.3 \pm 2.6$   & 
 $10.2-14.6$  \\
$K^{*\pm}\pi^{\mp}$ & 
 $16^{+6}_{-5} \pm 2 $ &
 $<30$ &  
 $-$ &
 $16 \pm 6$    &
 $8.0-11.6$    \\
$K^{*+} \pi^{0}$ & 
 $<31$ &
 $-$ &  
 $-$ &
 $<31$   & 
 $2.0-5.1 $  \\
$K^{*0}\pi^{0}$ & 
 $<3.6 $ &
 $<7$ &  
 $-$ &
 $<3.6$    &
 $1.8-4.4$  \\   
\hline 
\end{tabular}
%\end{center}
\end{table} 
%%%%%%%%%%%%%%%%%%%%%%%%%%%%%%%%%%%%%%%%%%%%%%%%%%%%%%%%%%%%%%%%%%%%%%%%%

%%%%%%%%%%%%%%%%%%%%%%% Ratios of CP-averaged rates  %%%%%%%%%%%%%%%%%%%%
\begin{table}[hb]
\begin{tabular}{|c||c||c|c|} \hline 
~~~~~~Quatity~~~~~~ & ~~~~Experiment~~~~ & ~~~~~~~~~PQCD~~~~~~~~~  
& ~~~~~~QCDF\cite{neub:a}~~~~~  \\ \hline 
  &   &   &   \\
${Br(\pi^{+} \pi^{-}) \over Br(\pi^{\pm} K^{\mp}) }$ & $0.25 \pm 0.04$ & 
 $0.30-0.69$ & $0.5-1.9$  \\
  &   &   &   \\
${Br(\pi^{\pm} K^{\mp}) \over 2 Br(\pi^{0} K^{0}) }$ & $1.05 \pm 0.27$ & 
 $0.78-1.05$ & $0.9-1.4$  \\
  &   &   &   \\
${2 \,\, Br(\pi^{0} K^{\pm}) \over Br(\pi^{\pm} K^{0}) }$ & $1.25 \pm 0.22$ & 
 $0.77-1.60$ & $0.9-1.3$  \\
  &   &   &   \\
${\tau(B^{+}) \over \tau(B^0)} \,
{Br(\pi^{\mp} K^{\pm}) \over Br(\pi^{\pm} K^{0}) }$ & $1.07 \pm 0.14$ & 
 $0.70-1.45$ & $0.6-1.0$  \\
  &   &   &   \\
\hline
\end{tabular}
\label{TABLE11.5}
\caption{Ratios of CP-averaged rates in $B \to K \pi, \pi\pi $ decays 
with $\phi_3=80^0$, $R_b=0.38$. Here we adopted
$m_0^{\pi}=1.3$ GeV and $m_0^{K}=1.7$ GeV.}
\end{table} 
%%%%%%%%%%%%%%%%%%%%%%%%%%%%%%%%%%%%%%%%%%%%%%%%%%%%%%%%%%%%%%%%%%%%%%%%%
%%%%%%%%%%%%%%%%%%%%%%% CP-Asymmety in Kpi, pipi decays  %%%%%%%%%%%%%%%%%%%%
\begin{table}[t]
\begin{tabular}{|c||c|c||c|c|} \hline 
~~Direct~~$A_{CP}(\%)$~~~~~~& ~~~~~BELLE~~~~~   & ~~~~~BABAR~~~~~   
& ~~~~~~~~~~PQCD~~~~~~~~~~ & ~~~~~~~QCDF~~~~~~ \\ \hline 
$\pi^{+}\pi^{-}$ & $58\pm15\pm7$ & $9 \pm 15 \pm 4$ & 
 $16.0 \sim 30.0$ & $-6\pm12$  \\ \hline
$\pi^{+}\pi^{0}$ & $-14\pm 24^{+5}_{-4}$ & $1 \pm 10 \pm 2$ & 
 $0.0$ & 0.0 \\ \hline 
$\pi^{0}\pi^{0}$ & $43 \pm 51^{+17}_{-16}$ & $12 \pm 56 \pm 6$ & 
 $20.0 \sim 40.0$ & $-$  \\ \hline \hline
$\pi^{+} K^{-}$ & $-10.1 \pm 2.5 \pm 0.5$ & $-13.3\pm 3.0 \pm 0.9$ & 
$-12.9 \sim -21.9  $ & $5\pm9$   \\ \hline
$\pi^{0}K^{-}$ & $4\pm 5\pm 2$ & $6.0 \pm 6.0 \pm 1.0$ & 
 $-10.0 \sim -17.3$ & $7\pm9$  \\ \hline
$\pi^{-}\bar{K}^{0}$ & $7^{+9+1}_{-8-3}$ & $-8.7 \pm 4.6 \pm 1.0$ & 
 $-0.6 \sim -1.5$ & $1\pm1$   \\ \hline 
$\pi^{0}{K}^{0}$ & $16 \pm 29 \pm 5$ & $-6 \pm 18 \pm 6$ & 
 $-0.90 \sim -1.03$ & $-3.6 \sim 0.8 $   \\ \hline 
\end{tabular}
\label{TABLE3}
\caption{CP-asymmetry in $B \to K \pi, \pi\pi $ decays 
with $\phi_3=40^0 \sim 90^0$, $R_b=\sqrt{\rho^2+\eta^2}=0.38$. 
Here we adopted $m_0^{\pi}=1.3$ GeV and $m_0^{K}=1.7$ GeV.}
\end{table} 
%%%%%%%%%%%%%%%%%%%%%%%%%%%%%%%%%%%%%%%%%%%%%%%%%%%%%%%%%%%%%%%%%%%%%%%%%
%%%%%%%%%%%%%%%%%%%%%%%%%%%%%%%%%%%%%%%%%%%%%%%%%%%%%%%%%%%%%%%%%%%%%%%%%
\vspace{3mm}
{\bf CP Asymmetry of $B \to \pi\pi, K\pi$:}
Because we have a large imaginary contribution from factorized 
annihilation diagrams in PQCD approach,
we predict large CP asymmetry ($\sim 25 \%$) in $B^0 \to \pi^{+}\pi^{-}$ decays
and about $-15 \%$ CP violation effects in  $B^0 \to K^{+}\pi^{-}$.
The detail prediction is given in Table ~6.
The CP asymmetry is defined as followings:
\begin{eqnarray}
A_{CP}(\triangle t) &=& 
{ N(\bar{B} \to \bar{f}) - N(B \to f) \over N(\bar{B} \to \bar{f}) + N(B \to f)} \nonumber \\
&=& S_{f}\,\, sin(\triangle m_d \triangle t) -C_{f}\,\, Cos(\triangle m_d \triangle t). 
\end{eqnarray} 
Here we notice that the relation between two different definitions: 
$A_f(Belle)=-C_f(BaBar)$. In our analysis we used the Belle notation.
The precise measurement of direct CP asymmetry (both magnitude and sign) 
is a crucial way to test factorization models 
which have different sources of strong phases.
Our predictions for CP-asymmetry on $B\to K(\pi)\pi$ have a totally opposite
sign to those of QCD factorization. Recently it was confirmed 
as the first evidence of the direct CP-violation in B-decays that the
DCP asymmetry in $B \to K^{\pm}\pi^{\mp}$ decay are $-10.1 \pm 2.6 \%$ with
$3.9\sigma$ deviations from zero in Belle Coll., and $-13.3\pm 3.1 \%$ with 
$4.2\sigma$, which is in a good agreement with PQCD result\cite{KLS}.

%%%%%%% Determination of Phi_2 and Phi_3 from B to pipi/Kpi %%%%%%%%%%%%%
\section{Extraction of $\phi_2(=\alpha)$ from $B \to \pi^{+}\pi^{-}$}
Even though isospin analysis of $B \to \pi\pi$ can provide a clean way
to determine $\phi_2$, it might be difficult in practice because of
the large uncertainty of the branching ratio of $B^0 \to \pi^0\pi^0$.
In reality in order to determine $\phi_2$, we can use the time-dependent rate
of $B^0(t) \to \pi^{+}\pi^{-}$.
Since penguin contributions are sizable about 20-30 \% of the total amplitude,
we expect that direct CP violation can be large if strong phases are different
in the tree and penguin diagrams.

In our analysis we use the c-convention.
The ratio between penguin and tree amplitudes is $R_c=|P_c/T_c|$ 
and the strong phase difference
between penguin and tree amplitudes $\delta=\delta_P-\delta_T$.
The time-dependent asymmetry measurement provides two equations for
$C_{\pi\pi}$ and $S_{\pi\pi}$ in terms of three unknown variables 
$R_c,\delta$ and $\phi_2$\cite{GR}.
%%%%%%%%%%%%%%%%%%%%%%%%%%%%
\begin{figure}
%\epsfxsize=6cm 
%\centerline{ angle=-90,\epsfbox{Acppipi-00.eps}}
\begin{center}
\includegraphics[angle=-90,width=0.5\textwidth]{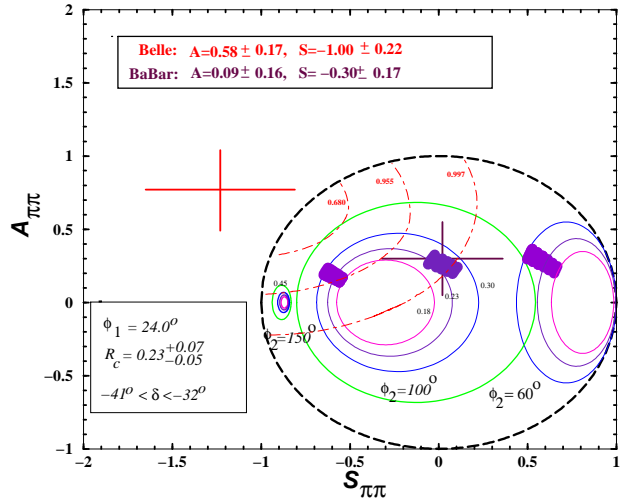} 
\caption{Plot of $A_{\pi\pi}$ versus $S_{\pi\pi}$  for various values
of $\phi_2$ with $\phi_1=24.3^o$, $0.18 < R_c < 0.30$ and $-41^o <
\delta < -32^o$ in the pQCD method.}
\end{center}
\label{fig:cpipi}
\end{figure}
%%%%%%%%%%%%%%%%%%%%%%%%%%%
\begin{figure}
\epsfxsize=8cm 
\centerline{\epsfbox{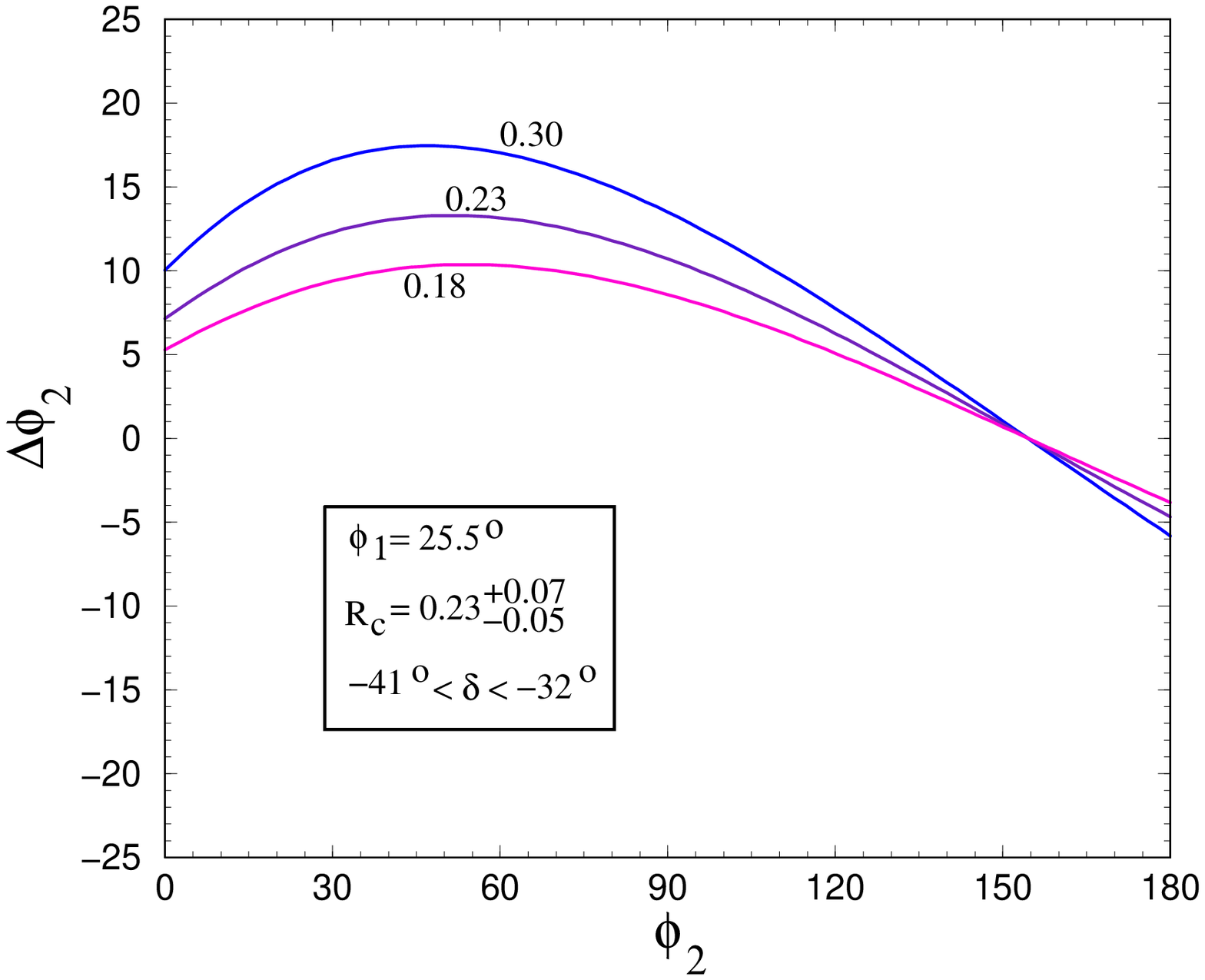}}
\begin{center}
\caption{Plot of $\Delta \phi_2$ versus $\phi_2$ with
$\phi_1=25.5^o$, $0.18 < R_c < 0.30$ and $-41^o <
\delta < -32^o$ in the PQCD method.}
\end{center}
\label{fig:delphi2}
\end{figure}
%%%%%%%%%%%%%%%%%%%%%%%%%%%
Since PQCD provides us $R_c=0.23^{+0.07}_{-0.05}$ and $-41^o
<\delta<-32^o$, the allowed range of $\phi_2$ at present stage is
determined as $55^o <\phi_2< 100^o$ as shown in Figure \ref{fig:cpipi}. 

According to the power counting rule in the PQCD approach,
the factorizable annihilation contribution with large imaginary part
becomes subdominant and give a negative strong phase from 
$-i\pi\delta(k_{\perp}^2-x\,M_B^2)$.
Therefore we have a relatively large
strong phase in contrast to the QCD-factorization ($\delta\sim 0^o$) 
and predict large direct CP violation effect 
in $B^0\to \pi^{+}\pi^{-}$ 
with $A_{cp}(B^0 \to \pi^{+}\pi^{-}) = (23\pm7) \%$, 
which will be tested by more precise experimental measurement within two years. 

In the numerical analysis, though the data by Belle
collaboration\cite{belle} 
is located ourside allowed physical regions, 
we considered the averaged value of recent  measurements\cite{belle,babar}:
\begin{itemize}
\item[$\bullet$]
$S_{\pi\pi}=  -0.30\pm0.17\pm0.03$ (BaBar), \hspace{8mm}
$S_{\pi\pi}=  -1.00\pm0.21\pm0.07$ (Belle); 
\item[$\bullet$]
$A_{\pi\pi}= 0.09\pm0.15\pm0.04$ (BaBar), \hspace{10mm}
$A_{\pi\pi}= 0.58\pm0.15\pm0.07$ (Belle). 
\end{itemize}
The central point of averaged data corresponds to $\phi_2 = 78^o$ 
in the PQCD method. 
Even if the data by Belle
collaboration\cite{belle} 
is located ourside allowed physical regions, we can have allowed
ranges with 2 $\sigma$ bounds, but large negative $\delta$ and 
$R_c > 0.4$ is prefered\cite{KS-ckm03}.
%%%%%%%%%%%%%%%%%%%%%%%%%%%%%%%%%%%%%%%%%%%%%%%%%%%%%%%%%%%%%%%%%%

\section{Extraction of $\phi_3(=\gamma)$ 
from $B^0 \to K^{+}\pi^{-}$ and $B^{+}\to K^0\pi^{+}$}
By using tree-penguin interference in $B^0\to K^{+}\pi^{-}(\sim
T^{'}+P^{'})$ versus $B^{+}\to K^0\pi^{+}(\sim P^{'})$, CP-averaged
$B\to K\pi$ branching fraction may lead to non-trivial constaints
on the $\phi_3$ angle\cite{fle-man}. In order to determine $\phi_3$,
we need one more useful information 
on CP-violating rate differences\cite{gr-rs02}.
Let's introduce the following observables :
\beqa
R_K &=&{\overline{Br}(B^0\to K^{+}\pi^{-}) \,\, \tau_{+} \over
\overline{Br}(B^+\to K^{0}\pi^{+}) \,\, \tau_{0} }
= 1 -2\,\, r_K \, cos\delta \, \, cos\phi_3 + r_K^2 \nonumber \\
&& \hspace{40mm} \geq sin^2\phi_3     \\
\cr
A_0 &=&{\Gamma(\bar{B}^0 \to K^{-}\pi^{+} - \Gamma(B^0 \to
K^{+}\pi^{-}) \over \Gamma(B^{-}\to \bar{K}^0\pi^{-}) +
 \Gamma(B^{+}\to \bar{K}^0\pi^{+}) } \nonumber \\
&=& A_{cp}(B^0 \to K^{+}\pi^{-}) \,\, R_K = -2 r_K \, sin\phi_3 \,sin\delta.
\eeqa
where $r_K = |T^{'}/P^{'}|$ is the ratio of tree to penguin amplitudes
and $\delta = \delta_{T'} -\delta_{P'}$ is the strong phase difference
between tree and penguin amplitides.
After eliminate $sin\delta$ in Eq.(8)-(9), we have
\beq
R_K = 1 + r_K^2 \pm \sqrt(4 r_K^2 cos^2\phi_3 -A_0^2 cot^2\phi_3).
\eeq
Here we obtain $r_K = 0.201\pm 0.037$ 
from the PQCD analysis\cite{KLS:02,keum} 
and $A_0=-0.11\pm 0.065$ by combining recent
measurements on CP asymmetry of $B^0\to K^+\pi^-$: 
$A_{cp}(B^0\to K^+\pi^-)=-11.7\pm2.8\pm0.7 \%$ \cite{babar,belle}
with present world averaged value of  $R_K=1.10\pm 0.15$\cite{rk}.

%%%%%%%%%%%%%%%%%%%%%%%%%%%
\begin{figure}[htbp]
%\epsfxsize=6cm
%\centerline{\epsfbox{Rkpi-fig.eps}}
\begin{center}
\includegraphics[angle=-90,width=0.6\textwidth]{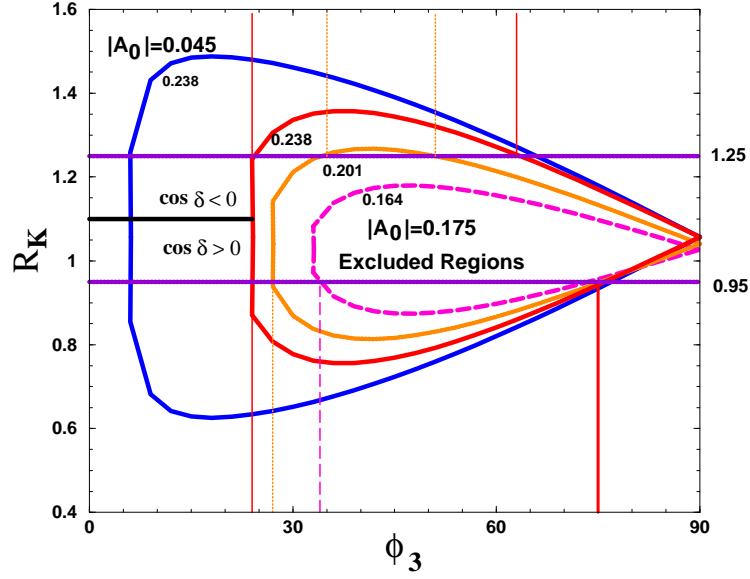} 
\caption{Plot of $R_K$ versus $\phi_3$ with $r_K=0.164,0.201$ and $0.238$.}
\end{center}
\label{fig:Rkpi}
\end{figure}
%%%%%%%%%%%%%%%%%%%%%%%%%%%
%%%%%%%%%%%%%%%%%%%%%%%%%%%
As shown in Fig.~12, we can constrain $\phi_3$ 
with $1\,\sigma$ range of World Averaged $R_K$ as follows:
\begin{itemize}
\item[$\bullet$]For $cos\delta > 0$, $r_K=0.164$: we can exclude
$0^o \leq \phi_3 \leq 6^0$ and $ 24^o \leq \phi_3 \leq 75^0$. 
\item[$\bullet$]For $cos\delta > 0$, $r_K=0.201$: we can exclude
$0^o \leq \phi_3 \leq 6^0$ and $ 27^o \leq \phi_3 \leq 75^0$. 
\item[$\bullet$]For $cos\delta > 0$, $r_K=0.238$: we can exclude
$0^o \leq \phi_3 \leq 6^0$ and $ 34^o \leq \phi_3 \leq 75^0$.
\item[$\bullet$]For $cos\delta < 0$, $r_K=0.164$: we can exclude
$0^o \leq \phi_3 \leq 6^0$. 
\item[$\bullet$]For $cos\delta < 0$, $r_K=0.201$: we can exclude
$0^o \leq \phi_3 \leq 6^0$ and $ 35^o \leq \phi_3 \leq 51^0$. 
\item[$\bullet$]For $cos\delta < 0$, $r_K=0.238$: we can exclude
$0^o \leq \phi_3 \leq 6^0$ and $ 24^o \leq \phi_3 \leq 62^0$.
\end{itemize}

According to the table 2, since
we obtain $\delta_{P'} = 157^o$ and $\delta_{T'} = 1.4^o$,
the value of $cos\delta$ becomes negative, $-0.91$.
Therefore the maximum value of the constraint bound for the $\phi_3$
is strongly depend on the uncertainty of $|V_{ub}|$.
When we take the central value of $r_K=0.201$,
$\phi_3$ is allowed within the ranges of $51^o \leq \phi_3 \leq
129^o$, which is consistent with the results by the model-independent
CKM-fit in the $(\rho,\eta)$ plane.
 
%%%%%%%%%%%%%%%%%%%%%%%%%%%%%%%%%%%%%%%%%%%%%%%%%%%%%%%%%%%%%%%%%%

%%%%%%%%%%%%%%%%%%%%%%%%%%%%%%%%%%%%%%%%%%%%%%%%%%%%%%%%%%%%%%%%%%
\vspace{5mm}
\section{Radiative B-decays ($B \to (K^*/\rho/\omega) \gamma$):} 
Radiative B-meson decays can provide the most reliable window to understand
the framework of the Standard Model(SM) and to look for New Physics beyond SM
by using the rich sample of B-decays.

In contrast to the inclusive radiative B-decays, exclusive processes such as
$B \to K^{*}\gamma$ are much easier to measure in the experiment with a good
precision\cite{Nakao03}. 

%%%%%%%%%%%%%%%%%%%%%%%%%%%%%%%%%%%%%%%%%%%%%%%%%%%%%%%%%%%%%%%%%
\begin{table}[htb]
\vspace*{0.5cm}
%\hspace*{-1.2cm}
%\begin{center}
\begin{tabular}{|c||c|c|c|}   \hline 
Decay Modes & ~~CLEO~~ & ~~BaBar~~ & ~~Belle~~ 
\\ \hline \hline
${\cal Br}(B \to K^{*0}\gamma $) ($10^{-5}$) &
$4.55 \pm 0.70 \pm 0.34$  & $4.23 \pm 0.40 \pm 0.22$ & 
$4.09 \pm 0.21 \pm 0.19$ \\ \hline
${\cal Br}(B \to K^{*\pm}\gamma $)($10^{-5}$) &
$3.76 \pm 0.86 \pm 0.28 $ & $3.83 \pm 0.62 \pm 0.22$ & 
$4.40 \pm 0.33 \pm 0.24 $ \\ \hline
${\cal Br}(B \to \rho^{0}\gamma $) ($10^{-6}$) &
$< 17$  & $<1.2$ & $< 2.6$ \\ \hline
${\cal Br}(B \to \rho^{+}\gamma $) ($10^{-6}$) &
$< 13 $ & $< 2.1$ & $< 2.7 $ \\ \hline
${\cal Br}(B \to \omega\gamma $) ($10^{-6}$) &
   & $< 1.0$ & $< 4.4 $ \\ \hline \hline
${\cal A}_{CP}(B \to K^{*0}\gamma )$ ($\%$) &
$8 \pm 13 \pm 3$  & $-3.5 \pm 9.4 \pm 2.2$ & 
$-6.1 \pm 5.9 \pm 1.8$  \\ \hline
${\cal A}_{CP}(B \to K^{*+}\gamma )$ ($\%$) &
  &  & $+5.3 \pm 8.3 \pm 1.6$ \\ \hline 
\end{tabular}
\caption{Experimental measurements of the averaged branching ratios and 
CP-violating asymmetries of the exclusive $B\to V\gamma$ decays for 
$V=K^{*},\rho$ and $\omega$.
\label{table1}}
%\end{center}
\end{table}
%%%%%%%%%%%%%%%%%%%%%%%%%%%%%%%%%%%%%%%%%%%%%%%%%%%%%%%%%%%%%%%%%%%%%%%%%%%%%%%%%%
%%%%%%%%%%%%%%%%%%%%%%%%%%%%
\begin{figure}[htbp]
\epsfxsize=10cm
\centerline{\epsfbox{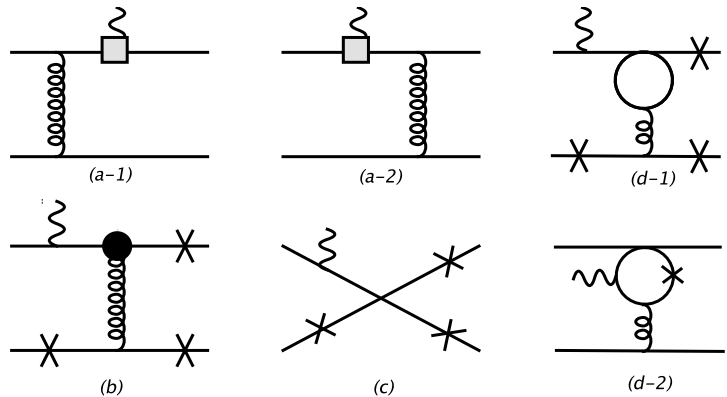}}
%\begin{center}
%\resizebox{25pc}{!} {\includegraphics[angle=0,width=10.0cm]{kstg.eps}} 
%\begin{center}
\caption{Feynman diagrams of the magnetic penguin(a), chromomagnetic penguin(b),
annihilation(c) and $0_2$-penguin contributions for $B \to V\gamma$ decays}
%\end{center}
\label{fig:kstg}
\end{figure}
%%%%%%%%%%%%%%%%%%%%%%%%%%%
The main short-distance (SD) contribution to the $B \to K^* \gamma$ decay rate 
involves the matrix element
\begin{equation}
<K^* \gamma| O_{7} |B> ={e m_b \over 8 \pi^2} (-2 i) \epsilon^{\mu}_{\gamma}
<K^* |\bar{s} \sigma_{\mu\nu}q^{\nu}(1-\gamma_5) b | B(p)>,
\end{equation}
which is parameterized in terms of two invariant form fectors as 
\begin{eqnarray}
<K^*(P_3,\epsilon_3)|\bar{s} \sigma_{\mu\nu}q^{\nu}(1-\gamma_5) b | B(P)>
&=& [\epsilon_{3,\mu}(q\cdot P)-P_{\mu}(q\cdot\epsilon_3)] \cdot 2 T_2(q^2) 
\nonumber \\
&& \hspace{15mm} +i\epsilon_{\mu\nu\alpha\beta} \epsilon_3^{\nu} P^{\alpha}
q^{\beta} \cdot 2 T_1(q^2).
\end{eqnarray}
Here $P$ and $P_3=P-q$ are the B-meson and $K^*$ meson momentum, respectively
and $\epsilon_3$ is the polarization vector of the $K^*$ meson.
For the real photon emission process the two form factors coincide,
$T_1(0)=T_2(0)=T(0)$. 
This form factor can be calculable in the $k_T$ factorization
method including the sudakov suppression factor and 
the threshold resummation effects. As discussed in ref\cite{KM03}, 
we obtain $T(0)=0.28 \pm 0.02 $ for $B \to K^* \gamma$ 
which is far away from 
the QCD result $0.38\pm 0.06$ by using the light-cone QCD sum rule \cite{BB-98}, 
however in accordance with the preliminary result of Lattice QCD, 
$0.25 \pm 0.06$\cite{Beci-03}.

Even though theoretical predictions for the exclusive decays always has
large model dependent hadronic uncertainties, such uncertainties can be 
cancelled in the searching of the CP-asymmetry and the isospin breaking effect.

Including all possible contributions from $0_{7\gamma},0_{8g},0_2$-penguin
and annihilation in Figure.~13, we obtain 
the Branching ratios\cite{KM03,ICHP03}:
\begin{itemize}
\item $Br(B^0 \to K^{0*}\gamma) = (3.5^{+1.1}_{-0.8} ) \times 10^{-5}$ \hspace{2mm};
\hspace{2mm} $Br(B^+ \to K^{+*}\gamma) = (3.4^{+1.2}_{-0.9}) \times 10^{-5}, $
\item $Br(B^0 \to \rho^{0}\gamma) = (0.95 \pm 0.14)\times 10^{-6}$ \hspace{1mm};
\hspace{1mm}$Br(B^+ \to \rho^{+}\gamma) = (1.63 \pm 0.40)\times 10^{-6}$, 
\end{itemize}
and the CP-Asymmetry :
\begin{itemize}
\item $Acp(B^0 \to K^{0*}\gamma) = (0.39^{+0.06}_{-0.07} ) \% $ \hspace{10mm}
$Acp(B^+ \to K^{+*}\gamma) = (0.62 \pm 0.13)  \%$
\end{itemize}
The small difference in the branching fraction between $K^{0*}\gamma$
and $K^{+*}\gamma$ can be detected as the isopsin symmetry breaking  
which tells us the sign of the combination of the Wilson coefficients, $C_6/c_7$.
We obtain
\begin{equation}
\Delta_{0-}={\eta_{\tau} Br(B \to \bar{K}^{0*}\gamma) - Br(B \to K^{*-}\gamma)
\over \eta_{\tau} Br(B \to \bar{K}^{0*}\gamma) + Br(B \to K^{*-}\gamma) }
= (5.7^{+ 1.1}_{-1.3}  \pm 0.8  ) \%
\end{equation}
where $\eta_{\tau}=\tau_{B^+}/\tau_{B^0}$. The first error term comes from
the uncertainty of shape parameter of the B-meson wave function 
($0.36 < \omega_B < 0.44$) in charm penguin contribution
and the second term is origined from the uncertainty of $\eta_{\tau}$. 
By using the world averaged value of measurement and 
$\tau_{B^+}/\tau_{B^0}=1.083 \pm 0.017$, we find numerically that
$\Delta_{0-}(K^*\gamma)^{exp}=(3.9 \pm 4.8) \%$. In PQCD
large isospin symmetry breaking in $B \to K^* \gamma$ system cannot be expected.

%%%%%%%%%%%%%%%%%%%%%%%%%%%%%%%%%%%%%%%%%%%%%%%%%%%%%%%%%%%%%%%%%

\section{Summary and Outlook}
In this paper I have summarized ingredients of $k_T$-factorization
approach and some important
theoretical predictions by comparing exparimental data,
which is based on my previous works\cite{KLS:01,KLS:02,KLS:03,KS-03}.
The PQCD factorization approach provides a useful theoretical framework
for a systematic analysis on non-leptonic two-body B-meson decays including
radiative decays. Our results are in a good agreement with experimental data. 
Specially PQCD predicted large direct CP asymmetries
in $B^0 \to \pi^{+}\pi^{-}, K^{+}\pi^{-}$ decays, 
which will be a crucial touch stone to distinguish our approach 
from others in future precise measurement. Recently the measurement
of the direct CP asymmetry in $B\to K^{\pm}\pi^{\mp}$, 
$A_{cp}(K^{+}\pi^{-})=-12\pm 3 \%$ is in accordance with our prediction.

We discussed the method to determine weak phases 
$\phi_2$ within the PQCD approach 
through Time-dependent asymmetries in $B^0\to
\pi^{+}\pi^{-}$. 
We get interesting bounds on $55^o < \phi_2 < 100^o$ with
90\% C.L. of the recent averaged measurements.
%%%%%%%%%%%%%%%%%%%%%%%%%%%%%%%%%%%%%%%%%%%%%%%%%%%%%%%%%%%%%%%%%%

\vskip1.5mm
{\bf Acknowledgments}
It is a great pleasure to thank D.P. Roy, A. Kundu and Uma Shanka 
for their hospitality at WHEPP8-workshop, Mumbai in India. 
I wish to acknowlege the fruitful collaboration 
and joyful discussions with other members of PQCD working group.
This work was supported in part by a visiting scholar program in DESY
and in part by Grant-in Aid from NSC: NSC 92-2811-M-001-088 in Taiwan.

\end{document}